\documentclass[conference, 10pt]{IEEEtran}

\usepackage{amsmath,amssymb,epsfig,multicol}
\usepackage{epsf}

\begin{document}

\title{\huge{Power-Bandwidth Tradeoff in Multiuser Relay Channels}\\[0.5ex]
       \huge{with Opportunistic Scheduling}\\[1.5ex]} 

\author{\authorblockN{\"Ozg\"ur Oyman}
\authorblockA{
Intel Research\\
Santa Clara, CA 95054\\
Email: ozgur.oyman@intel.com}\and
\authorblockN{Moe Z. Win}
\authorblockA{
Massachusetts Institute of Technology (MIT)\\
Cambridge, MA 02139\\
Email: moewin@mit.edu}
}


%


\maketitle

\begin{abstract}
The goal of this paper is to understand the key merits of multihop relaying techniques jointly in terms of their energy efficiency and spectral efficiency advantages in the presence of multiuser diversity gains from opportunistic (i.e., channel-aware) scheduling and identify the regimes and conditions in which relay-assisted multiuser communication provides a clear advantage over direct multiuser communication. For this purpose, we use Shannon-theoretic tools to analyze the tradeoff between energy efficiency and spectral efficiency (known as the power-bandwidth tradeoff) over a fading multiuser relay channel with $K$ users in the asymptotic regime of large (but finite) number of users (i.e., dense network). Benefiting from the extreme-value theoretic results of \cite{Oyman_isit07}, we characterize the power-bandwidth tradeoff and the associated energy and spectral efficiency measures of the bandwidth-limited high signal-to-noise ratio (SNR) and power-limited low SNR regimes, and utilize them in investigating the large system behavior of the multiuser relay channel as a function of the number of users and physical channel SNRs. Our analysis results in very accurate closed-form formulas in the large (but finite) $K$ regime that quantify energy and spectral efficiency performance, and provides insights on the impact of multihop relaying and multiuser diversity techniques on the power-bandwidth tradeoff. 
\end{abstract}



%
\IEEEpeerreviewmaketitle

\section{Introduction} 

We consider the uplink and downlink of a cellular multihop/mesh system (e.g., IEEE 802.16j systems), with one base station, one fixed relay station and $K$ users. The role of the relay station is to enhance end-to-end link quality in terms of capacity, coverage and reliability using multihop routing techniques \cite{Oyman_mag07}, and its presence allows the base station to choose between (i) sending/receiving data directly to/from a given user, (ii) communicating over a two-hop route where the base station sends data to the relay station and the relay station forwards the data to the users in downlink, and vice versa for the uplink. We refer to this communication model as the {\it multiuser relay channel}; which includes both the multiaccess relay channel (uplink) \cite{Kramer05, Laneman_ciss} and broadcast relay channel (downlink) \cite{Kramer05, Liang07}. 

For fixed portable applications, where radio channels are slowly varying, multiple access methods based on opportunistic scheduling mechanisms take advantage of variations in users' channel quality and allocate resources such that the user with the best channel quality is served at any given time or frequency (could be subject to certain fairness and delay constraints). It has been shown by the pioneering works \cite{Knopp95}-\nocite{Hanly98}\cite{Viswanath02} that the sum capacity under such opportunistic scheduling algorithms increases with the number of users; yielding {\it multiuser diversity} gains by exploiting the time and frequency selectivity of wireless channels as well as the independent channel variations across users. 

{\it Relation to Previous Work:} While multiuser diversity concepts over traditional cellular systems is well understood, there are open issues on the design and analysis of resource allocation and opportunistic scheduling algorithms in relay-based cellular multihop networks. Recently, low-complexity resource management methods for OFDMA-based cellular multihop networks were proposed in \cite{Oyman_asil06}-\cite{Oyman_globecom07}, and were shown to simultaneously realize gains from both multiuser diversity and multihop relaying to enhance capacity and coverage. Moreover, in \cite{Oyman_isit07}, tools from extreme-value theory were used to characterize the spectral efficiency of opportunistic scheduling algorithms over fading multiuser relay channels in the asymptotic regime of large (but finite) number of users, and to provide insights on the role of multiuser diversity, multihop routing and spectrum reuse techniques in leveraging the system-level performance. Finally, applications of opportunistic communication principles to relay-assisted wireless networks were investigated in various other contexts in \cite{Win07}\nocite{Gunduz07,Lo07}-\cite{Zhang04}.

{\it Contributions:} The prior art on the analysis of opportunistic scheduling algorithms over the multiuser relay channel has focused on spectral efficiency, which is clearly an important performance metric given the scarcity of bandwidth resources in a cellular network and system operation in the bandwidth-limited regime. Often however, each user's mobile terminal in a cellular network could additionally be severely constrained by its computational and transmission/receiving power and/or could be subject to poor signal quality due to path loss and shadow fading effects of the wireless channel (e.g., mobile terminal in a coverage hole at the cell edge), leading to system operation in the power-limited regime, where the optimization of energy efficiency becomes crucial in system design.

The goal of this paper is to understand the key merits of multihop relaying techniques jointly in terms of their energy efficiency and spectral efficiency advantages in the presence of multiuser diversity gains from opportunistic (i.e., channel-aware) scheduling and identify the regimes and conditions in which relay-assisted multiuser communication provides a clear advantage over direct multiuser communication. For this purpose, we use Shannon-theoretic tools to analyze the tradeoff between energy efficiency and spectral efficiency (known as the power-bandwidth tradeoff) over a fading multiuser relay channel with $K$ users in the asymptotic regime of large (but finite) number of users (i.e., dense network). Benefiting from the extreme-value theoretic results of \cite{Oyman_isit07}, we characterize the power-bandwidth tradeoff and the associated performance measures of the high and low signal-to-noise ratio (SNR) regimes, and utilize them in investigating the large system behavior of the multiuser relay channel as a function of the number of users and physical channel SNRs. Our analysis results in very accurate formulas in the large (but finite) $K$ regime, and provides insights on the impact of multihop relaying and multiuser diversity techniques on the power-bandwidth tradeoff.

\section{Multiuser Relay Channel Model and Protocol Assumptions}

\begin{figure}[t]  

 \centering

  \includegraphics[height=!,width=3in]{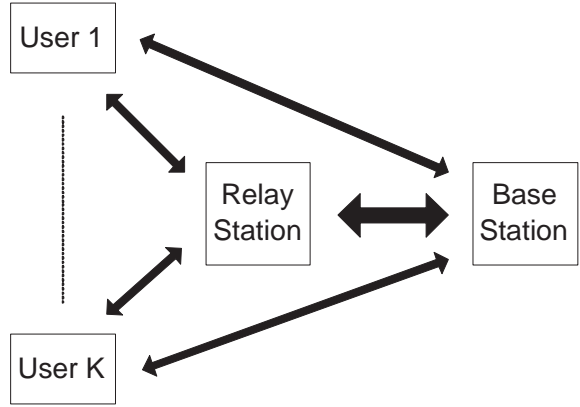}

  \caption{Multiuser relay channel model (both uplink and downlink).}

  \label{micro_net}

\end{figure}

Consider the network depicted in Fig. \ref{micro_net} with $K+2$ nodes, in which $K$ users, indexed by $k=1,...,K$, send/receive information to/from a base station. The relay station is designated to help users transmit/receive information utilizing its high capacity backhaul link to the base station. In other words, multiple users share a single relay for uplink and downlink (i.e. multiaccess and broadcast). 

We assume that all the links over the multiuser relay channel are corrupted by additive white Gaussian noise (AWGN). Furthermore, the links between the base station and users are assumed to be under frequency-flat multiplicative fading i.i.d. across users, with complex channel gains $\{h_k\}_{k=1}^K$, where $|h_k|^2 \in {\Bbb R}$ is a real-valued random variable representing the channel power gain for the link between the base station and user $k$, drawn from an arbitrary continuous distribution $F_h$ with ${\Bbb E}[|h_k|^2]=1,\,\forall k$. The average received SNR between the base station and each user equals $\mathsf{SNR}^{(b)}$. Analogously, the links between the relay station and users have average SNRs given by $\mathsf{SNR}^{(r)}$,\footnote{Superscript $b$ stands for "base" station and $r$ stands for "relay" station.} and are under frequency-flat fading i.i.d. across users, with complex channel gains $\{g_k\}_{k=1}^K$, where $|g_k|^2 \in {\Bbb R}$ represents the channel power gain for the link between the relay station and user $k$, drawn from an arbitrary continuous distribution $F_g$ with ${\Bbb E}[|g_k|^2]=1,\,\forall k$. The set of channels $\{h_k\}_{k=1}^K$ and $\{g_k\}_{k=1}^K$ are independent. It is assumed that the cellular backhaul link between the base station and relay station is an AWGN line-of-sight (LOS) connection with received SNR equal to $\mathsf{SNR}_{B}$. 

Furthermore, our channel model concentrates on the quasi-static regime, in which, once drawn, the channel gains $\{h_k\}_{k=1}^K$ and $\{g_k\}_{k=1}^K$ over the multiuser links remain fixed for the entire duration of a time slot allocated for codeword transmission i.e., each codeword spans a single fading state, and that the channel coherence time is much larger than the coding block length, i.e., slow fading assumption. Due to slow fading, each terminal in the multiuser relay channel is able to obtain full channel state information regarding its transmission/reception links. Because the transmission rate of each codeword over each hop is adaptively chosen so that reliable decoding is always possible (the data rate is changed on a codeword by codeword basis to adapt to the instantaneous channel fading conditions), system is never in outage provided that the coding blocklength is arbitrarily large. 

For the following analysis, we assume a time-division based (half duplex) relaying constraint for multi-hop routing protocols, which is due to the practical limitation that terminals can often not transmit and receive at the same time. In particular, we consider a two-phased decode-and-forward protocol; where, for a given routing path between the base station and a given user, the relay station hears and fully decodes the transmitted data signal in the first phase and forwards its re-encoded version in the second phase. Moreover, we consider the use of point-to-point capacity achieving codes only (without any kind of cooperation across terminals) over all transmissions over the multiuser relay channel and do not impose any delay constraints on the multihop system; in contrast, we allow each
coded transmission to have an arbitrarily large blocklength, although we will be concerned with the relative sizes of blocklengths over multiple hops.

\section{Asymptotic Measures of Energy and Spectral Efficiency}

This section describes our methodology for evaluating power-bandwidth tradeoff in multiuser relay channels and accordingly introduces the key measures of energy and spectral efficiency to be used in our performance characterization. As part of the prior art in this respect, the analytical tools to study the power-bandwidth tradeoff in the power-limited regime have been previously developed in the context of point-to-point
single-user communications \cite{Verdu02}-\cite{Lozano03}, and were extended to multi-user (point-to-multipoint and multipoint-to-point) settings
\cite{Shamai01}-\nocite{Caire04}\nocite{Lapidoth03}\cite{Muharem03}, as well as to adhoc wireless networking examples of
single-relay channels \cite{Abbas03}-\cite{Yao03}, AWGN multihop networks \cite{Laneman05}-\nocite{Oyman06b}\cite{Oyman_isssta08} and dense multi-antenna relay networks \cite{oyman_pbt2005}. In the bandwidth-limited regime, the necessary tools to perform the
power-bandwidth tradeoff analysis were developed by \cite{Shamai01} in the
context of code-division multiple access (CDMA) systems and were
later used by \cite{Lozano05} and \cite{Jindal05} to characterize fundamental limits
in multi-antenna channels over point-to-point and broadcast communication, respectively, and by \cite{Oyman06b}, \cite{oyman_pbt2005} to study a variety of adhoc networking scenarios.
 
We assume that all terminals in the multiuser relay channel are supplied with finite total average transmit power $P$ (in Watts (W)) over unconstrained bandwidth $B$ (in Hertz (Hz)). The multiuser relay channel with desired end-to-end data rate $R$ must respect the fundamental limit $R/B \leq {\mathsf C}\left(E_b/N_0\right)$,
where $\mathsf{C}$ is the Shannon capacity (ergodic mutual information\footnote{We emphasize that due to our assumptions on the channel statistics stated in Section II, a Shannon capacity exists (this is obtained by averaging the total mutual information over the statistics of the channel processes) for the multiuser relay channel.} ) (in bits/second/Hertz or b/s/Hz), which we will also refer as the spectral efficiency, and $E_b/N_0$ is the energy per information bit normalized by background noise spectral level, expressed as
$E_b/N_0 = \mathsf{SNR}/C(\mathsf{SNR})$, for $\mathsf{SNR} = P/(N_0B)$ and $C$ denoting the spectral efficiency as a function of $\mathsf{SNR}$ \footnote{The use of $C$ and $\mathsf{C}$ avoids assigning the same symbol to spectral efficiency functions of $\mathsf{SNR}$
and ${E_b}/{N_0}$.} (in nats/second/Hertz or nats/s/Hz).
There exists a tradeoff between the efficiency measures ${E_b}/{N_0}$ and ${\mathsf C}$ (known as the power-bandwidth tradeoff)
in achieving a given target data rate. When
${\mathsf C} \ll 1$, the system operates in the {\it power-limited
regime}; i.e., the bandwidth is large and the main concern is the
limitation on power. Similarly, the case of ${\mathsf C} \gg 1$
corresponds to the {\it bandwidth-limited regime}. Tightly framing achievable performance, particular emphasis in our
power-bandwidth tradeoff analysis is placed on the regions of low and high ${E_b}/{N_0}$.

{\it Low ${E_b}/{N_0}$ regime:} Defining
$({E_b}/{N_0})_{\mathrm{min}}$ as the minimum system-wide
${E_b}/{N_0}$ required to convey any positive rate reliably, we have
$(E_b/N_0)_{\mathrm{min}} = \min \, \mathsf{SNR}/C(\mathsf{SNR})$, 
over all $\mathsf{SNR} \geq 0$. In most scenarios, ${E_b}/{N_0}$ is minimized
in the power-limited wideband regime when $\mathsf{SNR}$ is low and $C$ is near zero.
We consider the first-order behavior of $\mathsf{C}$ as a function of
${E_b}/{N_0}$ when $\mathsf{C} \rightarrow 0$ by analyzing the affine function (in decibels)
\footnote{ $\,\,u(x)=o(v(x)), x \rightarrow L$ stands for $\lim_{x
\rightarrow L}\frac{u(x)}{v(x)}=0$.} 
$$
10\log_{10}\frac{E_b}{N_0} \left( {\mathsf C} \right) =
10\log_{10}\frac{E_b}{N_0}_{\mathrm{min}} +
\frac{\mathsf C}{S_0}10\log_{10}2 + o({\mathsf C}),
$$
where $S_0$ denotes the ``wideband'' slope of spectral
efficiency in b/s/Hz/(3 dB) at the point $({E_b}/{N_0})_{\mathrm{min}}$, 
$$
S_0 = \lim_{\frac{E_b}{N_0} \downarrow \frac{E_b}{N_0}_{\mathrm{min}}}
\frac{{\mathsf C}(\frac{E_b}{N_0})}{10\log_{10}\frac{E_b}{N_0}-10\log_{10}
\frac{E_b}{N_0}_{\mathrm{min}}}10\log_{10}2 .
$$
It can be shown that \cite{Verdu02}
\begin{equation}
\frac{E_b}{N_0}_{\mathrm{min}} = \lim_{\mathsf{SNR} \rightarrow 0} \frac{\ln{2}}{\dot{C}(\mathsf{SNR})},
\label{wideband_slope1}
\end{equation}
and
\begin{equation}
S_0 = \lim_{\mathsf{SNR} \rightarrow 0} \frac{2{\left[ \dot{C}(\mathsf{SNR}) \right]}^2}{-\ddot{C}(\mathsf{SNR})},
\label{wideband_slope}
\end{equation}
where $\dot{C}$ and $\ddot{C}$ denote the first and second order derivatives of
$C(\mathsf{SNR})$ (evaluated in nats/s/Hz). 

{\it High ${E_b}/{N_0}$ regime:} In the high SNR regime (i.e., $\mathsf{SNR} \rightarrow \infty$), the
dependence between ${E_b}/{N_0}$ and $\mathsf{C}$ can be characterized as \cite{Shamai01}
\begin{eqnarray}
10\log_{10} \frac{E_b}{N_0} ({\mathsf C}) & = & \frac{\mathsf C}{S_{\infty}} 10 \log_{10} 2 - 10 \log_{10}({\mathsf C}) \nonumber \\
& + & 10 \log_{10} \frac{E_b}{N_0}_{\mathrm{imp}} + o(1), \nonumber
\end{eqnarray}
where $S_{\infty}$ denotes the ``high SNR'' slope of the spectral efficiency in b/s/Hz/(3 dB)
\begin{eqnarray}
S_{\infty} & = & \lim_{\frac{E_b}{N_0} \rightarrow \infty} \frac{{\mathsf C}(\frac{E_b}{N_0})}{10\log_{10} 
\frac{E_b}{N_0}} 10\log_{10} 2 \nonumber \\
& = & \lim_{\mathsf{SNR} \rightarrow \infty} \mathsf{SNR} \, \dot{C}(\mathsf{SNR})  \label{spec_slope} 
\end{eqnarray}  
and $({E_b}/{N_0})_{\mathrm{imp}}$ is the ${E_b}/{N_0}$ improvement factor with
respect to a single-user single-antenna unfaded AWGN reference
channel\footnote{For the AWGN channel; $C(\mathsf{SNR}) =
\ln(1+\mathsf{SNR})$ resulting in $S_0 = 2$,
$({E_b}/{N_0})_{\mathrm{min}} = \ln{2}$, $S_{\infty} = 1$ and $({E_b}/{N_0})_{\mathrm{imp}} = 1$.} and it is expressed as
\begin{equation}
\frac{E_b}{N_0}_{\mathrm{imp}} = \lim_{\mathsf{SNR} \rightarrow \infty}
\left[ \, \mathsf{SNR} \, 
\exp \left( \, - \, \frac{C(\mathsf{SNR})}{S_{\infty}} \, \right)
\right].
\label{pwr_offset}
\end{equation}

\section{Power-Bandwidth Tradeoff Characterization}

Consider the scheduling problem such that $K$ users in the multiuser relay channel are to be assigned time-slots for transmission/reception over a common bandwidth. This problem involves transmissions over three types of links: (i) $L_B$: Wireless cellular backhaul link between the base station and relay station with received SNR equal to $\mathsf{SNR}_{B}$, (ii) $L_R$: The link between the relay station and users with average SNR equal to $\mathsf{SNR}^{(r)}$ and complex channel gains $\{g_k\}_{k=1}^K$ and (iii) $L_D$: The direct link between the base station and users with average SNR equal to $\mathsf{SNR}^{(b)}$ and complex channel gains $\{h_k\}_{k=1}^K$. In this context, we employ the maximum signal-to-interference-plus-noise ratio (max-SINR) opportunistic scheduling algorithm, which always serves the user with the best instantaneous rate at any given time/frequency resource, in conjunction with two transmission protocols over the multiuser relay channel:

{\it a) Direct transmission (no relaying):} Only link $L_D$ is active for all available time resources. The base station compares the instantaneous rates over the direct links between itself and the users determined by the channel gains $\{h_k\}_{k=1}^K$ and assigns link $L_D$ to the best user with the highest instantaneous rate.

{\it b) Two-hop relaying:} Links $L_B$ and $L_R$ are active for this relay-assisted multihop routing protocol. We assign positive time-sharing coefficients $\beta_B \in [0,1]$ and $\beta_R \in [0,1]$ to links $L_B$ and $L_R$, respectively, to specify the fractional time during which these links are active, such that $\beta_B + \beta_R = 1$. The relay station compares the instantaneous rates over the links between itself and the users determined by the channel gains $\{g_k\}_{k=1}^K$ and and assigns link $L_R$ to the best user with the highest instantaneous rate. The base station communicates with the selected user over a two-hop route through the relay station. 

\subsection{Opportunistic Multiuser Scheduling with Direct Communication}

Assuming Gaussian inputs, i.e., all input signals have the temporally i.i.d. zero-mean circularly symmetric complex Gaussian distribution, the maximum supportable spectral efficiency over the multiuser relay channel achieved by direct communication in conjunction with opportunistic scheduling is given by (in nats/s/Hz) 
\begin{eqnarray}
\mathsf{C}^{\mathrm{direct}}(\mathsf{SNR}) & = & {\Bbb E} \left[ \max_{k=1,...,K} \, C(\mathsf{SNR} \, \alpha^{(b)} \, |h_k|^2) \right]  \nonumber \\ && \label{spec_eff_direct} \\
& = & c_K^{(h)}(\mathsf{SNR}) \, \kappa + d_K^{(h)}(\mathsf{SNR}), \label{speceff_direct_closed}
\end{eqnarray}
where the closed form expression in (\ref{speceff_direct_closed}) follows from direct application of the extreme-value theoretic analysis in \cite{Oyman_isit07} for the asymptotic regime of large $K$ and assumes Type I convergence \cite{Leadbetter83} on the maxima of the i.i.d. instantaneous spectral efficiency realizations in (\ref{spec_eff_direct}), that depend on the channel power gains $\{|h_k|^2\}_{k=1}^K$, i.e., $F_h$ belongs to Type I domain of attraction leading to the Gumbel limiting extreme-value distribution. In (\ref{spec_eff_direct}), the spectral efficiency function $C(x)$ is defined as $C(x)=\ln(1+x)$ and we express $\mathsf{SNR}^{(b)}$ as $\mathsf{SNR}^{(b)} = \alpha^{(b)} \, \mathsf{SNR}$. Moreover, in (\ref{speceff_direct_closed}), $\kappa \approx 0.57721566$ is Euler's constant and $c_K^{(h)}(\mathsf{SNR})$ and $d_K^{(h)}(\mathsf{SNR})$ are given by 
$$
c_K^{(h)} (\mathsf{SNR}) = \frac{\mathsf{SNR} \, \alpha^{(b)} \, a_K^{(h)}}{1 + \mathsf{SNR} \, \alpha^{(b)} \, b_K^{(h)}}, 
$$
$$
d_K^{(h)} (\mathsf{SNR}) = \ln (1+\mathsf{SNR} \, \alpha^{(b)} \, b_K^{(h)}),
$$
where $a_K^{(h)}$ and $b_K^{(h)}$ are the sequences of constants necessary to ensure the following convergence in distribution as $K \rightarrow \infty$ (cf. Lemma 1 in \cite{Oyman_isit07}): 
\begin{equation}
{\Bbb P} \left( \frac{\max_k \, |h_k|^2 - b_K^{(h)}}{a_K^{(h)}} \leq x \right) \rightarrow \exp(-\exp(-x)).
\label{cond_max1}
\end{equation}
The following theorem states our results on spectral efficiency vs. $E_b/N_0$ characterization for direct communication in the presence of multiuser diversity gains from opportunistic scheduling: 
\\

{\bf \noindent Theorem 1:} {\it In the asymptotic regime of large $K$, assuming that $F_h$ belongs to Type I domain of attraction leading to the Gumbel limiting extreme-value distribution, the power-bandwidth tradeoff for direct communication with opportunistic scheduling can be characterized through the following relationships: }
\\

\underline{\it Low $\frac{E_b}{N_0}$ regime:}
$$\,\,\,\,
\frac{E_b}{N_0}_{\mathrm{min}}^{\mathrm{direct}}=\frac{\ln{2}}{\alpha^{(b)}\,\left(\kappa\,a_K^{(h)} + b_K^{(h)}\right)},  
$$
$$\mathrm{and}
\,\,\,\,\,\,\,\, S_0^{\mathrm{direct}} = \frac{2\,\left(\kappa\,a_K^{(h)}+b_K^{(h)}\right)^2}{b_K^{(h)}\,\left(2\,\kappa\,a_K^{(h)} + b_K^{(h)}\right)}. 
$$

\underline{\it High $\frac{E_b}{N_0}$ regime:}
$$\,\,\,\,
\frac{E_b}{N_0}_{\mathrm{imp}}^{\mathrm{direct}}=
\frac{1}{\alpha^{(b)}\,b_K^{(h)}} \exp \left(-\frac{a_K^{(h)}}{b_K^{(h)}}\,\kappa \right), 
$$
$$\mathrm{and}
\,\,\,\,\,\,\,\, S_{\infty}^{\mathrm{direct}} = 1.
$$
We clearly see from Theorem 1 how multiuser diversity gains impact the energy and spectral efficiency measures of the bandwidth-limited high SNR and power-limited low SNR regimes. In particular, the scaling constants $a_K^{(h)}$ and $b_K^{(h)}$ impact both of $(E_b/N_0)_{\mathrm{min}}^{\mathrm{direct}}$ and $(E_b/N_0)_{\mathrm{imp}}^{\mathrm{direct}}$, as well as $S_0^{\mathrm{direct}}$, although it should be noted that $S_0$ converges to $2$ as $K \rightarrow \infty$. Assuming Rayleigh fading distribution on $F_h$, we have $a_K^{(h)} = 1$ and $b_K^{(h)} = \log(K)$, which implies that $(E_b/N_0)_{\mathrm{min}}^{\mathrm{direct}}$ and $(E_b/N_0)_{\mathrm{imp}}^{\mathrm{direct}}$ decay at a rate of at least $1/\log(K)$ in the regime of asymptotically large $K$.

\subsection{Opportunistic Multiuser Scheduling with Two-Hop Relaying}

Assuming Gaussian inputs, the maximum supportable end-to-end spectral efficiency over the multiuser relay channel achieved by two-hop relaying in conjunction with opportunistic scheduling is given by (in nats/s/Hz) 
\begin{eqnarray}
\mathsf{C}^{\mathrm{relay}} & = & \min \left[ \beta_B \, C\left(\mathsf{SNR}\right), \right. \nonumber \\ 
& & \left. \beta_R \, \max_{k=1,...,K} \, C \left( \mathsf{SNR} \, \alpha^{(r)} \, |g_k|^2 \right) \right]
\nonumber \\
& & \label{spec_eff_relay} \\
& = & \beta_B \, C(\mathsf{SNR}) - \beta_R \, c_K^{(g)}(\mathsf{SNR}) \, \mathrm{Ei}(z_{K}(\mathsf{SNR})), \nonumber \\
&& \label{speceff_relay_closed} 
\end{eqnarray}
where the closed form expression in (\ref{speceff_relay_closed}) follows from the extreme-value theoretic analysis in \cite{Oyman_isit07} for the asymptotic regime of large $K$ (cf. Theorem 2 in \cite{Oyman_isit07}) and assumes Type I convergence on the maxima of the i.i.d. instantaneous spectral efficiency realizations in (\ref{spec_eff_relay}), that depend on the channel power gains $\{|g_k|^2\}_{k=1}^K$, i.e., $F_g$ belongs to Type I domain of attraction leading to the Gumbel limiting extreme-value distribution. In (\ref{spec_eff_relay}), we express $\mathsf{SNR}^{(r)}$ and $\mathsf{SNR}_B$ as $\mathsf{SNR}^{(r)} = \alpha^{(r)} \, \mathsf{SNR}$ and $\mathsf{SNR}_B = \mathsf{SNR}$. Moreover, in (\ref{speceff_relay_closed}), $\mathrm{Ei}(x)$ is the exponential integral function defined by $\mathrm{Ei}(x)=\int_x^{\infty} \frac{e^{-y}}{y} \,dy$, $z_{K}(\mathsf{SNR})$ is given by 
$$
z_{K}(\mathsf{SNR}) = \exp\left(\frac{\beta_R\,d_K^{(g)}(\mathsf{SNR})-\beta_B\,C(\mathsf{SNR}_B)}{\beta_R\,c_K^{(g)}(\mathsf{SNR})}\right).
$$ 
and $c_K^{(g)}(\mathsf{SNR})$ and $d_K^{(g)}(\mathsf{SNR})$ are given by 
$$
c_K^{(g)} (\mathsf{SNR}) = \frac{\mathsf{SNR} \, \alpha^{(r)} \, a_K^{(g)}}{1 + \mathsf{SNR} \, \alpha^{(r)} \, b_K^{(g)}}, 
$$
$$
d_K^{(g)} (\mathsf{SNR}) = \ln (1+\mathsf{SNR} \, \alpha^{(r)} \, b_K^{(g)}),
$$
where $a_K^{(g)}$ and $b_K^{(g)}$ are the sequences of constants necessary to ensure the following convergence in distribution as $K \rightarrow \infty$ (cf. Lemma 1 in \cite{Oyman_isit07}): 
\begin{equation}
{\Bbb P} \left( \frac{\max_k \, |g_k|^2 - b_K^{(g)}}{a_K^{(g)}} \leq x \right) \rightarrow \exp(-\exp(-x)).
\label{cond_max2}
\end{equation}
The following theorem states our results on spectral efficiency vs. $E_b/N_0$ characterization for two-hop relaying in the presence of multiuser diversity gains from opportunistic scheduling: 
\\

{\bf \noindent Theorem 2:} {\it In the asymptotic regime of large $K$, assuming that $F_g$ belongs to Type I domain of attraction leading to the Gumbel limiting extreme-value distribution, the power-bandwidth tradeoff for two-hop relaying with opportunistic scheduling can be characterized through the following relationships: }
\\

\underline{\it Low $\frac{E_b}{N_0}$ regime:}
$$
\frac{E_b}{N_0}_{\mathrm{min}}^{\mathrm{relay}}=\frac{\ln{2}}{\beta_B - \beta_R\,\alpha^{(r)}\,a_K^{(g)} \, \mathrm{Ei} \left( \zeta_K \right)},  
$$
$$\mathrm{and}
\,\,\,\,\,\,\,\, S_0^{\mathrm{relay}} = \frac{2\,\left( \beta_B - \beta_R \, a_K^{(g)} \, \alpha^{(r)} \, \mathrm{Ei} \left( \zeta_K \right) \right)^2}{\beta_B - 2\,\beta_R \, a_K^{(g)} \, b_K^{(g)} \, (\alpha^{(r)})^2 \, \mathrm{Ei} \left( \zeta_K \right)},   
$$
{\it where}
$$
\zeta_K = \exp\left( \frac{ \beta_R \, \alpha^{(r)} \, b_K^{(g)} - \beta_B}{\beta_R\,\alpha^{(r)}\,a_K^{(g)}} \right).
$$

\underline{\it High $\frac{E_b}{N_0}$ regime:}
$$
\frac{E_b}{N_0}_{\mathrm{imp}}^{\mathrm{relay}} = \left\{
\begin{array}{cc}
\left(\,1\,/\,\left(\alpha^{(r)}\,b_K^{(g)}\right)\right)\, \exp \left(-\,a_K^{(g)}\,\kappa \, / \, b_K^{(g)} \right) \\
\,\,\,\,\,\,\,\,\,\,\,\,\,\,\,\,\,\,\,\,\,\,\,\,\,\,\,\,\,\,\,\,\,\,\,\,\,\,\,\,\,\,\,\,\,\,\,\,\,\,\,\,\,\,\,\,\,\,\,\,\,\,\,\,\,\,\,\,\,\,\,\,\mathrm{for}\,\,\,\,\,\,\,\,\beta_B < \beta_R \\ \\
\exp\left(\left(a_K^{(g)}/b_K^{(g)}\right)\mathrm{Ei}\left( \left( \alpha^{(r)}b_K^{(g)} \right)^{b_K^{(g)}/a_K^{(g)}}\right)\right) \\
\,\,\,\,\,\,\,\,\,\,\,\,\,\,\,\,\,\,\,\,\,\,\,\,\,\,\,\,\,\,\,\,\,\,\,\,\,\,\,\,\,\,\,\,\,\,\,\,\,\,\,\,\,\mathrm{for} \,\,\,\,\,\,\,\,\beta_B = \beta_R = 1/2\\ \\
\,\,\,\,\,\,\,\,\,\,\,\,\,\,\,\,\,\,\,\,\,\,\,\,\,\,\,\,\,\,\,\,\,\, 1 \,\,\,\,\,\,\,\,\,\,\,\,\,\,\,\,\,\,\,\,\,\,\,\,\,\,\,\,\,\,\,\mathrm{for} \,\,\,\,\,\,\,\,\beta_B > \beta_R
\end{array}
\right.
$$
\\
$$\,\,\,\,\,\,\,\, \mathrm{and}
\,\,\,\,\,\,\,\,\,\,\,\,\,\,\,\, S_{\infty}^{\mathrm{relay}} = \min \left[\beta_B,\,\beta_R\right].
$$
\\
We clearly see from Theorem 2 how multiuser diversity gains impact the energy and spectral efficiency measures of the bandwidth-limited high SNR and power-limited low SNR regimes. In particular, the scaling constants $a_K^{(g)}$ and $b_K^{(g)}$ impact both of $(E_b/N_0)_{\mathrm{min}}^{\mathrm{relay}}$ and $(E_b/N_0)_{\mathrm{imp}}^{\mathrm{relay}}$, as well as $S_0^{\mathrm{relay}}$. Moreover, our analysis specifies the dependence of all power-bandwidth tradeoff measures on time-sharing coefficients $\beta_B$ and $\beta_R$. Finally, we note that the behavior of the energy efficiency measure of the bandwidth-limited high SNR regime, $(E_b/N_0)_{\mathrm{imp}}^{\mathrm{relay}}$, varies in a piecewise fashion as a function of $\beta_B$ and $\beta_R$.

\section{Numerical Results}

For the following numerical study, we assume Rayleigh fading distribution on $F_h$ and $F_g$ that leads to $a_K^{(h)} = a_K^{(g)} = 1$ and $b_K^{(h)} = b_K^{(g)} = \log(K)$, and furthermore set $K=20$,  $\alpha^{(b)}=0.01$, and $\alpha^{(r)}=1$. Selecting $K=20$ is typical as the number of users per sector in a relay-based cellular network (e.g., wireless metropolitan area networks (WMANs) based on the IEEE 802.16j multihop relay standard \cite{ieee802_16j}), and moreover the choices of $\alpha^{(b)}=0.01$, and $\alpha^{(r)}=1$ are realistic when the users are located at the cell edge, when the use of multihop relaying is most beneficial for leveraging the system-level performance of cellular networks \cite{Oyman_globecom07}. 

\begin{figure}  

 \centering

  \includegraphics[height=!,width=3.3in]{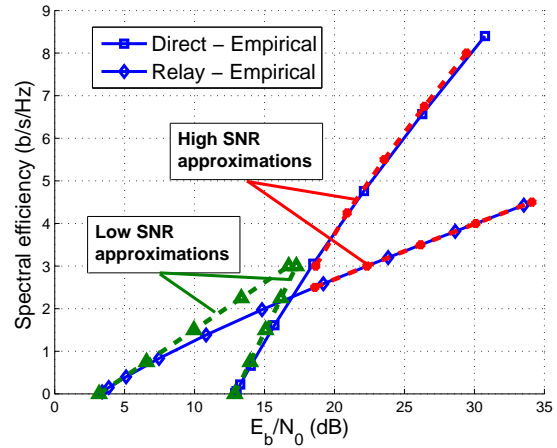}

  \caption{Spectral efficiency vs. $E_b/N_0$ for direct communication vs. two-hop relaying with opportunistic scheduling for  $\beta_B=1/3$ and $\beta_R=2/3$. Solid curves represent empirical power-bandwidth tradeoffs while dashed curves are analytical power-bandwidth tradeoffs.}

  \label{pbt}

\end{figure}

\begin{figure}  

 \centering

  \includegraphics[height=!,width=3.3in]{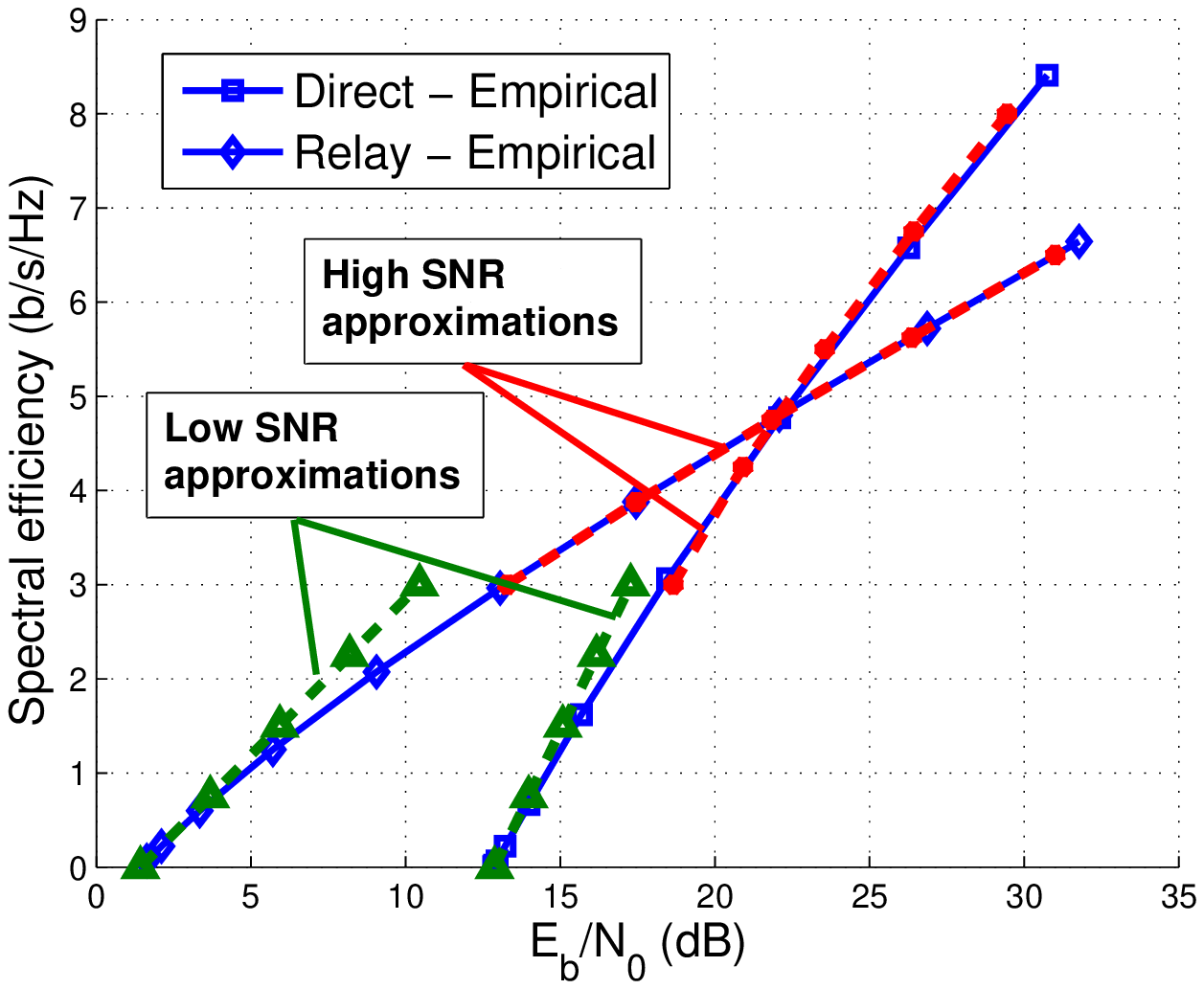}

  \caption{Spectral efficiency vs. $E_b/N_0$ for direct communication vs. two-hop relaying with opportunistic scheduling for  $\beta_B=1/2$ and $\beta_R=1/2$. Solid curves represent empirical power-bandwidth tradeoffs while dashed curves are analytical power-bandwidth tradeoffs.}

  \label{pbt2}

\end{figure}

\begin{figure}  

 \centering

  \includegraphics[height=!,width=3.3in]{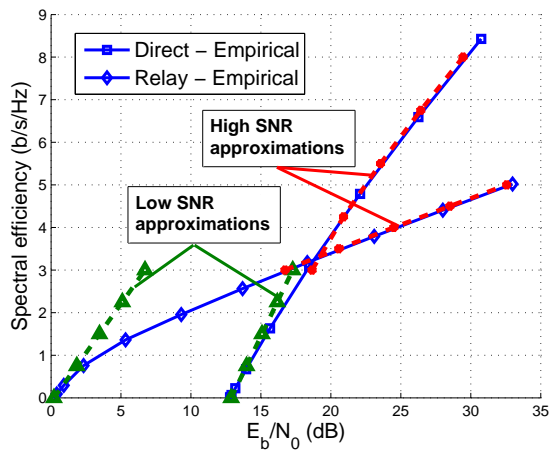}

  \caption{Spectral efficiency vs. $E_b/N_0$ for direct communication vs. two-hop relaying with opportunistic scheduling for  $\beta_B=2/3$ and $\beta_R=1/3$. Solid curves represent empirical power-bandwidth tradeoffs while dashed curves are analytical power-bandwidth tradeoffs.}

  \label{pbt3}

\end{figure}

In Figs. \ref{pbt}-\ref{pbt3}, we plot spectral efficiency vs. ${E_b}/{N_0}$ for direct communication and two-hop relaying in the presence of opportunistic scheduling for different values of $\beta_B$ and $\beta_R$. Here, we compare empirically generated power-bandwidth tradeoff curves (solid curves) with their analytical counterparts (dashed curves) from Theorems 1 and 2 for the low and high ${E_b}/N_0$ regimes. The empirical results are obtained by averaging the expressions in (\ref{spec_eff_direct}) and (\ref{spec_eff_relay}) over a large number of randomly generated fading realizations (based on Monte Carlo simulations) for various SNRs and computing the power-bandwidth tradeoffs from this set of average spectral efficiencies based on $E_b/N_0 = \mathsf{SNR}/C(\mathsf{SNR})$, where $C(\mathsf{SNR})$ is determined empirically. 

From Figs. \ref{pbt}-\ref{pbt3}, we validate the accuracy and tightness of the closed-form expressions presented in Theorems 1 and 2 for the power-bandwidth tradeoff in the scenarios of direct communication and two-hop relaying. In particular, we verify that our analytical results are well in agreement with the empirical results for all ranges of $\beta_B$ and $\beta_R$, despite the fact that $K$ is set at the finite value of $20$. Moreover, we observe that in the power-limited low SNR regime, multihop relaying provides a superior power-bandwidth tradeoff (and a significant power efficiency gain) over direct communication in the presence of multiuser diversity gains from opportunistic scheduling.

\begin{footnotesize}
\renewcommand{\baselinestretch}{0.2}
\bibliographystyle{IEEE}

\begin{thebibliography}{10}

\bibitem{Oyman_isit07}
\"O. Oyman,
\newblock ``Opportunism in multiuser relay channels: {S}cheduling, routing and
  spectrum reuse,''
\newblock in {\em Proc. IEEE International Symposium on Information Theory
  (ISIT'07)}, Nice, France, June 2007, pp. 286 -- 290.

\bibitem{Oyman_mag07}
\"O. Oyman, J.~N. Laneman, and S.~Sandhu,
\newblock ``Multihop relaying for broadband wireless mesh networks: {F}rom
  theory to practice,''
\newblock {\em IEEE Communications Magazine}, vol. 45, no. 11, pp. 116--122,
  Nov. 2007.

\bibitem{Kramer05}
G.~Kramer, M.~Gastpar, and P.~Gupta,
\newblock ``Cooperative strategies and capacity theorems for relay networks,''
\newblock {\em IEEE Trans. Inf. Theory}, vol. 51, no. 9, pp. 3037--3063, Sep.
  2005.

\bibitem{Laneman_ciss}
D.~Chen and J.~N. Laneman,
\newblock ``The diversity-multiplexing tradeoff for the multiaccess relay
  channel,''
\newblock in {\em Proc. Conference on Information Sciences and Systems (CISS)},
  Princeton, NJ, March 2006.

\bibitem{Liang07}
Y.~Liang and V.~V. Veeravalli,
\newblock ``Cooperative relay broadcast channels,''
\newblock {\em IEEE Trans. Inf. Theory}, vol. 53, no. 3, pp. 900--928, Mar.
  2007.

\bibitem{Knopp95}
R.~Knopp and P.~Humblet,
\newblock ``Information capacity and power control in single cell multiuser
  communications,''
\newblock in {\em Proc. IEEE Int. Computer Conf.}, Seattle, WA, Jun. 1995.

\bibitem{Hanly98}
D.~Tse and S.~Hanly,
\newblock ``Multi-access fading channels: Polymatroid structure, optimal
  resource allocation and throughput capacities,''
\newblock {\em IEEE Trans. Inf. Theory}, vol. 44, no. 7, pp. 2796--2815, Nov.
  1998.

\bibitem{Viswanath02}
P.~Viswanath, D.~N.~C. Tse, and R.~Laroia,
\newblock ``Opportunistic beamforming using dumb antennas,''
\newblock {\em IEEE Trans. Inf. Theory}, vol. 48, no. 6, pp. 1277--1294, Jun.
  2002.

\bibitem{Oyman_asil06}
\"O. Oyman,
\newblock ``{$\mathrm{OFDM^2A}$}: {A} centralized resource allocation policy
  for cellular multi-hop networks,''
\newblock in {\em Proc. IEEE Asilomar Conference on Signals, Systems and
  Computers}, Pacific Grove, CA, Oct. 2006.

\bibitem{Oyman_globecom07}
\"O. Oyman,
\newblock ``Opportunistic scheduling and spectrum reuse in relay-based cellular
  {OFDMA} networks,''
\newblock in {\em Proc. IEEE Global Telecommunications Conference
  (GLOBECOM'07)}, Washington, D.C., Nov. 2007, pp. 3699 -- 3703.

\bibitem{Win07}
A.~Bletsas, H.~Shin, and M.~Z. Win,
\newblock ``Cooperative communications with outage-optimal opportunistic
  relaying,''
\newblock {\em IEEE Trans. Wireless Communications}, vol. 6, no. 9, pp.
  3450--3460, Sep. 2007.

\bibitem{Gunduz07}
D.~G\"und\"uz and E.~Erkip,
\newblock ``Opportunistic cooperation by dynamic resource allocation,''
\newblock {\em IEEE Trans. Wireless Communications}, vol. 6, no. 4, pp.
  1446--1454, Apr. 2007.

\bibitem{Lo07}
C.~K. Lo, Jr. R.~W.~Heath, and S.~Vishwanath,
\newblock ``Opportunistic relay selection with limited feedback,''
\newblock in {\em Proc. IEEE VTC}, Dublin, Ireland, Apr. 2007, vol.~1, pp.
  135--139.

\bibitem{Zhang04}
M.~Hu and J.~Zhang,
\newblock ``Opportunistic multi-access: Multiuser diversity, relay-aided
  opportunistic scheduling, and traffic-aided smooth admission control,''
\newblock {\em Journal on Mobile Networks and Applications}, vol. 9, no. 4, pp.
  435--444, Aug. 2004.

\bibitem{Verdu02}
S.~Verd\'u,
\newblock ``Spectral efficiency in the wideband regime,''
\newblock {\em IEEE Trans. Inf. Theory}, vol. 48, no. 6, pp. 1319--1343, Jun.
  2002.

\bibitem{Lozano03}
A.~Lozano, A.~Tulino, and S.~Verd\'u,
\newblock ``Multiple-antenna capacity in the low-power regime,''
\newblock {\em IEEE Trans. Inf. Theory}, vol. 49, no. 10, pp. 2527--2543, Oct.
  2003.

\bibitem{Shamai01}
S.~Shamai (Shitz) and S.~Verd\'u,
\newblock ``The impact of flat-fading on the spectral efficiency of {CDMA},''
\newblock {\em IEEE Trans. Inf. Theory}, vol. 47, no. 5, pp. 1302--1327, May
  2001.

\bibitem{Caire04}
G.~Caire, D.~Tuninetti, and S.~Verd\'u,
\newblock ``Suboptimality of {TDMA} in the low-power regime,''
\newblock {\em IEEE Trans. Inf. Theory}, vol. 50, no. 4, pp. 608--620, Apr.
  2004.

\bibitem{Lapidoth03}
A.~Lapidoth, I.~E. Telatar, and R.~Urbanke,
\newblock ``On wide-band broadcast channels,''
\newblock {\em IEEE Trans. Inf. Theory}, vol. 49, no. 12, pp. 3250--3258, Dec.
  2003.

\bibitem{Muharem03}
T.~Muharemovi\'c and B.~Aazhang,
\newblock ``Robust slope region for wideband {CDMA} with multiple antennas,''
\newblock in {\em Proc. 2003 IEEE Information Theory Workshop}, Paris, France,
  March 2003, pp. 26--29.

\bibitem{Abbas03}
A.~El Gamal and S.~Zahedi,
\newblock ``Minimum energy communication over a relay channel,''
\newblock in {\em Proc. 2003 IEEE International Symposium on Information Theory
  (ISIT'03)}, Yokohama, Japan, June-July 2003, p. 344.

\bibitem{Yao03}
X.~Cai, Y.~Yao, and G.~Giannakis,
\newblock ``Achievable rates in low-power relay links over fading channels,''
\newblock {\em IEEE Trans. Communications}, vol. 53, no. 1, pp. 184--194, Jan.
  2005.

\bibitem{Laneman05}
M.~Sikora, J.~N. Laneman, M.~Haenggi, D.~J. Costello, and T.~E. Fuja,
\newblock ``Bandwidth and power efficient routing in linear wireless
  networks,''
\newblock {\em IEEE Trans. Inf. Theory}, vol. 52, no. 6, pp. 2624--2633, June
  2006.

\bibitem{Oyman06b}
\"O. Oyman and S.~Sandhu,
\newblock ``Non-ergodic power-bandwidth tradeoff in linear multi-hop
  networks,''
\newblock in {\em Proc. IEEE International Symposium on Information Theory
  (ISIT'06)}, Seattle, WA, July 2006, pp. 1514--1518.

\bibitem{Oyman_isssta08}
\"O. Oyman and J.~N. Laneman,
\newblock ``Multihop diversity in wideband {OFDM} systems: {T}he impact of
  spatial reuse and frequency selectivity,''
\newblock in {\em Proc. 2008 IEEE International Symposium on Spread Spectrum
  Techniques and Applications (ISSSTA'08)}, Bologna, Italy, Aug. 2008.

\bibitem{oyman_pbt2005}
\"O. Oyman and A.~J. Paulraj,
\newblock ``Power-bandwidth tradeoff in dense multi-antenna relay networks,''
\newblock {\em IEEE Trans. Wireless Communications}, vol. 6, no. 6, pp.
  2282--2293, June 2007.

\bibitem{Lozano05}
A.~Lozano, A.~Tulino, and S.~Verd\'u,
\newblock ``High-{SNR} power offset in multiantenna communication,''
\newblock {\em IEEE Trans. Inf. Theory}, vol. 51, no. 12, pp. 4134--4151, Dec.
  2005.

\bibitem{Jindal05}
N.~Jindal,
\newblock ``High {SNR} analysis of {MIMO} broadcast channels,''
\newblock in {\em Proc. 2005 IEEE International Symposium on Information Theory
  (ISIT'05)}, Adelaide, Australia, Sep. 2005.

\bibitem{Leadbetter83}
M.~R. Leadbetter, G.~Lindgren, and H.~Rootzen,
\newblock {\em Extremes and Related Properties of Random Sequences and
  Processes},
\newblock Springer-Verlag, New York, NY, 1983.

\bibitem{ieee802_16j}
{\em {IEEE} 802.16j: Relay Task Group Fixed Broadband Wireless Access},
\newblock URL: http://wirelessman.org/relay/index.html.

\end{thebibliography}

\end{footnotesize}

\end{document}